# A NOVEL APPROACH FOR PASS WORD AUTHENTICATION USING BIDIRECTIONAL ASSOCIATIVE MEMORY

*A S N Chakravarthy*[†] , *Penmetsa V Krishna Raja* [††] , Prof. *P S Avadhani*[†††]
[†] Associate Professor , Department of CSE& IT, Sri Aditya Engineering College, Surampalem, E.G.Dist , Andhra Pradesh, India
asnchakravarthy@yahoo.com
[††]Research Scholar, JNTUK, Kakinada, E.G.Dist , Andhra Pradesh, India
vamsilovesindia@gmail.com
[†††]Professor, Dept. of CS & SE, Andhra University, Visakhapatnam Dist, Andhra Pradesh, India
psavadhani@yahoo.com

## ABSTRACT

*Password authentication is a very important system security procedure to gain access to user resources. In the Traditional password authentication methods a server has check the authenticity of the users. In our proposed method users can freely select their passwords from a predefined character set. They can also use a graphical image as password. The password may be a character or an image it will be converted into binary form and the binary values will be normalized. Associative memories have been used recently for password authentication in order to overcome drawbacks of the traditional password authentication methods. In this paper we proposed a method using Bidirectional Associative Memory algorithm for both alphanumeric (Text) and graphical password. By doing so the amount of security what we provide for the user can be enhanced. This paper along with test results show that converting user password in to Probabilistic values and giving them as input for BAM improves the security of the system*

## 1. INTRODUCTION

The password authentication using HPNN takes the input and the output pattern together to the network. There is a chance of wrong user validation. In order to eliminate the limitations in PAS using HPNN in this paper introduces new technique i.e. password authentication using bidirectional associative memory.

## 2. BIDIRECTIONAL ASSOCIATIVE MEMORY (BAM):

The Bidirectional associative memory is heteroassociative, content-addressable memory. A BAM consists of neurons arranged in two layers say A and B. The neurons are bipolar binary. The neurons in one layer are fully interconnected to the neurons in the second layer. There is no interconnection among neurons in the same layer. The weight from layer A to layer B is same as the weights from layer B to layer A. dynamics involves two layers of interaction. Because the memory process information in time and involves Bidirectional data flow, it differs in principle from a linear association, although both networks are used to store association pairs. It also differs from the recurrent auto associative memory in its update mode[1]. The network structure of the Bi-directional Associative Memory model[2,3] is similar to that of the linear associator model, but the connections are bidirectional in nature, i.e., $w_{ij} = w_{ji}$, for i = 1, 2, ..., m and j = 1, 2, ..., n. The units in both layers serve as both input and output units depending on the direction of propagation. Propagating signals from the X layer to the Y layer makes the units in the X layer act as input units while the units in the Y layer act as output units. The same is true for the other direction, i.e., propagating from the Y layer to the X layer makes the units in the Y layer act as input units while the units in the X layer act as output units. Below is an illustration of the BAM architecture [4].

Just like the linear associator and Hopfield model, encoding in BAM can be carried out by using: to store a single associated pattern pair and

$$W_k = {X_k}^T Y_k \qquad (1)$$

$$W = \alpha \sum_{k=1}^{p} W_k \qquad (2)$$

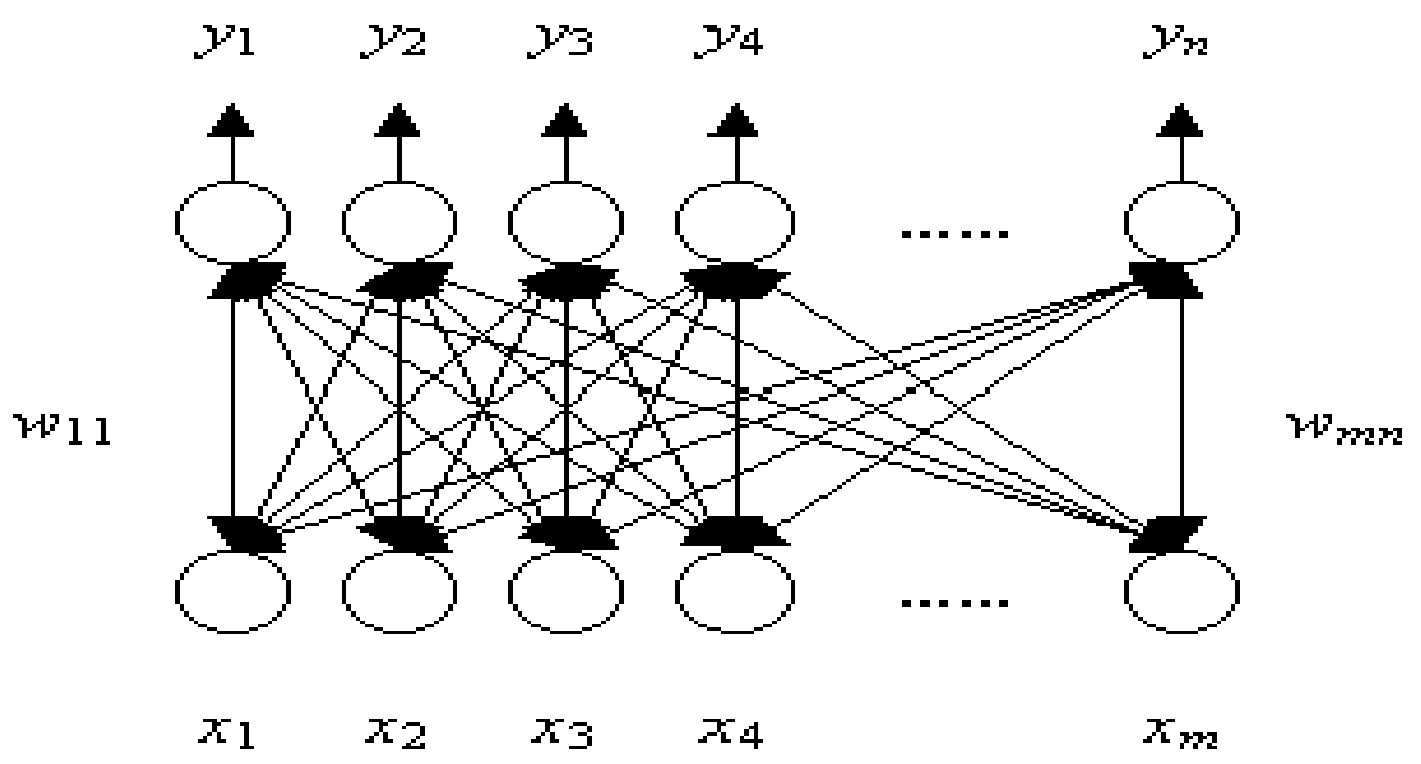

Figure 1. BAM Architecture

In BAM, decoding involves repeating the distributed information several times between the two layers until the network becomes stable. In decoding phase, an input pattern can be applied either on the X layer or on the Y layer. When an input pattern is given, the network will propagate the input pattern to the other layer allowing the units in the other layer to compute their output values. The pattern that was produced by the other layer is then propagated back to the original layer and let the units in the original layer compute their output values. The new pattern that was produced by the original layer is again propagated to the other layer. This process is repeated until further propagations and computations do not result in a change in the states of the units in both layers where the final pattern pair is one of the stored associated pattern pairs. [5, 6]

## 3. AUTHENTICATION USING BIDIRECTIONAL ASSOCIATIVE MEMORY:

The Bidirectional associative memory (BAM) is hetero associative, content-addressable memory. A BAM consists of neurons arranged in two layers say A and B. The neurons are bipolar binary. The neurons in one layer are fully interconnected to the neurons in the second layer. There is no interconnection among neurons in the same layer. The weight from layer A to layer B is same as the weights from layer B to layer A. dynamics involves two layers of interaction. Because the memory process information in time and involves Bidirectional data flow, it contradicts in principle from a linear association, although both networks are used to store association pairs. It also differs from the recurrent auto associative memory in its update mode. [7, 8]

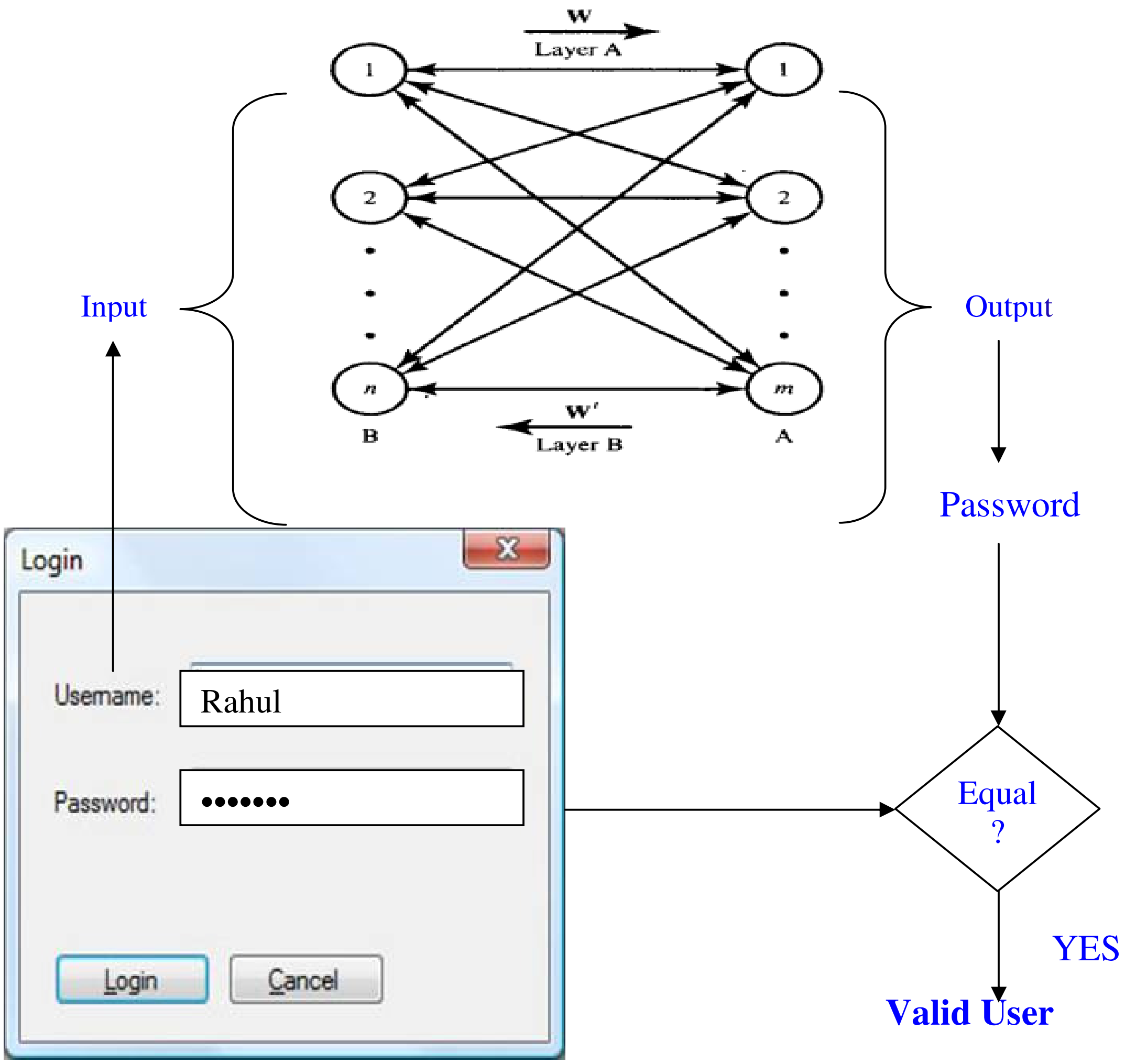

Figure 2 User Validation Using BAM

## 3.1 AUTHENTICATION PROCESS:

This method can use either textual or graphical password as input , so that it can authenticate the legitimate users.

## TEXT PASSWORD:

First this method converts the username and password into binary values and the uses those values as training samples, which can be performed by the following steps

- Convert each character into a unique number (for example ASCII value)

- Convert the unique number into binary value

A

↓

65 (ASCII value)

↓

01000001 (Binary Equivalent of 65)

Figure 3 Converting Character in to Binary Values

This procedure converts all the characters in the username and password into binary values.

Table 1 Binary Value for the Given User Name

| **Username** | **Binary value representing username** |
|---|---|
| RAJESH | 001001010100000100101001010100010110010100001001 |
| SWAMY | 0110010101110101010000010101100101001101 |
| KIRAN | 0110100101001001001001010100000100111001 |

After converting username and password into binary equivalents the pairs can be used as training samples. Once the training has been completed very soon the network will be stored in each server. When the user wants to use an application from a server he connects to the server and enters his username and password in that application, then server loads the BAM network and generates output by giving username as input. If the output matches with the password submitted by the user then server allows working with that application. The application can provide better authentication by using bipolar input instead of binary input. Application converts a binary number into bipolar number by using following formula or by simply replacing zeros with -1s. If Z is a binary digit then corresponding bipolar value is (2Z-1). [10]

1 → 1

0 → -1

The above procedure will reinforce in converting binary value in to bipolar value and can be used it as input to the network.

Table 2. Bipolar Values for the Given User Name

| **Username** | **Binary value representing username** |
|---|---|
| RAJESH | -1-11-1-11-11-11-1-1-1-1-11-1-11-11-1-11-11-11-1-1-11-111-1-11-11-1-1-1-11-1-11 |
| SWAMY | -111-1-11-11-1111-11-11-11-1-1-1-1-11-11-111-1-11-11-1-111-11 |
| KIRAN | -111-11-1-11-11-1-11-1-11-1-11-1-11-11-11-1-1-1-1-11-1-1111-1-11 |

## 3.2 LEARNING:

When a new user is interested to create an account, the network has to adjust weights so that it can recognize all the registered users with that application. The process of changing weights is in the network is as called learning.

## LEARNING IN BIDIRECTIONAL ASSOCIATIVE MEMORY:

Suppose we wish to store the binary (bipolar) patterns ($A_1$, $B_1$)… ($A_m$, $B_m$) at or near local energy minima. How can these association pairs be encoded in some BAM n-by-p matrix M. The association ($A_i$, $B_i$) can be viewed as a meta-rule or set-level logical implication: IF $A_i$ THEN $B_i$ However, bidirectionality implies that ($A_i$, $B_i$) also represents the converse meta-rule: IF Bi THEN Ai. Hence the logical relation between $A_i$ and $B_i$ is symmetric, namely, logical implication (set equivalence). The vector analogue of this symmetric biconditionality is correlation. The natural suggestion then is to memorize the association ($A_i$, $B_i$) by forming the correlation matrix or vector outer product $AT_iB_i$. The correlation matrix redundantly distributes the vector information in ($A_i$, $B_i$) in a parallel storage medium, a matrix. The next suggestion is to superimpose the m associations ($A_i$, $B_i$) by simply adding up the correlation matrices point wise. [ 11,12]

$$M = \sum_i A_i^T B_i \qquad (3)$$

With dual BAM memory MT given by

$$M^T = \sum_i (A_i^T B_i)^T = \sum_i B_i^T A_i \qquad (4)$$

## 3.3 BAM IMPLEMENTATION:

int No Of Patterns, No Of Bits Per Input, No Of Bits Per Output;

int[,] Weight, input, output, test;

Here *No Of Patterns* Specifies the number of patterns we want to use in the network training, *No Of Bits Per Input d*etermines number of bits we want to give for each input, *No Of Bits Per Output* Specifies number of bits we want to give for each output, *Weight Stores* specifies the weight values of the network, input Stores *input* vector, *output* Stores output vector and test stores pattern used for testing the BAM network.[10]

## 3.4 IMPLEMENTING TRAINING:

Before training starts the application will receive training samples from the user and stores them in the corresponding variables.

## THE BASIC TRAINING PROCEDURE

Consider N training pairs { ($A_1$, $B_1$), ($A_2$, $B_2$),…………,($A_l$ , $B_l$),……..,($A_N$ , $B_N$) where $A_l$ = ($a_{i1}$,$a_{i2}$ ,…….,$a_{im}$) and $B_l$ = ($b_{i1}$,$b_{i2}$ ,…….,$b_{im}$) and $a_{ij}$ , $b_{ij}$ are either in ON or OFF state. Where in binary mode, ON = 1 and OFF = 0 and in bipolar mode, ON = 1 and OFF = -1 The original weight matrix of the BAM is $X_i$

$$M_0 = \sum_{i=0}^{N} [X_i]^T [Y_i] \qquad (5)$$

where $X_i = (X_{i1}, X_{i2}, \ldots\ldots, X_{im})$ and $Y_i = (Y_{i1}, Y_{i2}, \ldots\ldots, Y_{ip})$
and $X(Y_{ij})$ is the bipolar form of $a_{ij}(b_{ij})$

```
private void Train()
{
   Weight = new int[input.GetLength(1), output.GetLength(1)];
   for(int i=0;i<input.GetLength(0);i++)
   {
      for (int  j= 0; j < input.GetLength(1); j++)
      {
         for (int k = 0; k < output.GetLength(1); k++)
              Weight[j, k] += input[i, j] * output[i, k];
 }
   }
}
```

**NOTE:** Here instead of calculating the transpose we multiply the required elements , which will be multiplied when we caluculate transpose.

## 3.5 RECOGNIZING THE PATTERN USING BAM:

The pattern which we want to use for testing the network will be supplied as input to the application and then application stores the pattern in the corresponding variable.

## THE METHODS AND THE EQUATIONS FOR RETRIEVE ARE:

Start with an initial condition which can be any given pattern pair (α , β).Determine a finite sequence of pattern pairs $(\alpha^1, \beta^1)$, $(\alpha^{11}, \beta^{11})$, until an equilibrium point $(\alpha_f, \beta_f)$ is reached, where

$B = \phi(A M)$ and $A^1 = \phi(B^1 M^T)$ (6)

$B^{11} = \phi(A^1 M)$ and $A^{11} = \phi(B^{11} M^T)$ (7)

$\Phi(F) = G = g_1, g_2, \ldots, g_r$ (8)

$F = (f_1, f_2, \ldots\ldots, f_r)$ (9)

M is correlation matrix

$$g_i = \begin{cases} 1 & \text{if } f_i > 0 \\ \begin{cases} 0 \text{ (binary)} \\ -1 \text{ (bipolar)} \end{cases} & , \ f_i < 0 \\ \text{Prevous } g_i & , \ f_i = 0 \end{cases} \qquad (10)$$

## PATTERN RECOGNIZATION IMPLEMENTATION USING BAM:

```
private int[,] Recognize()
{
                // pa -->   Previous alpha
  int[,] a =null, pb = null, b = null ;
  a = test;
  do{
    pb = Phi(MatrixMul(a, Weight));
    a = Phi(MatrixMul(pb, Transpose(Weight)));
    b = Phi(MatrixMul(a, Weight));
  }while(!areEqual(pb,b));
  ShowMatrix(pb);
  return pb;
}
```

## 4 RESULTS:

## BAM FOR TEXTUAL PASSWORDS:

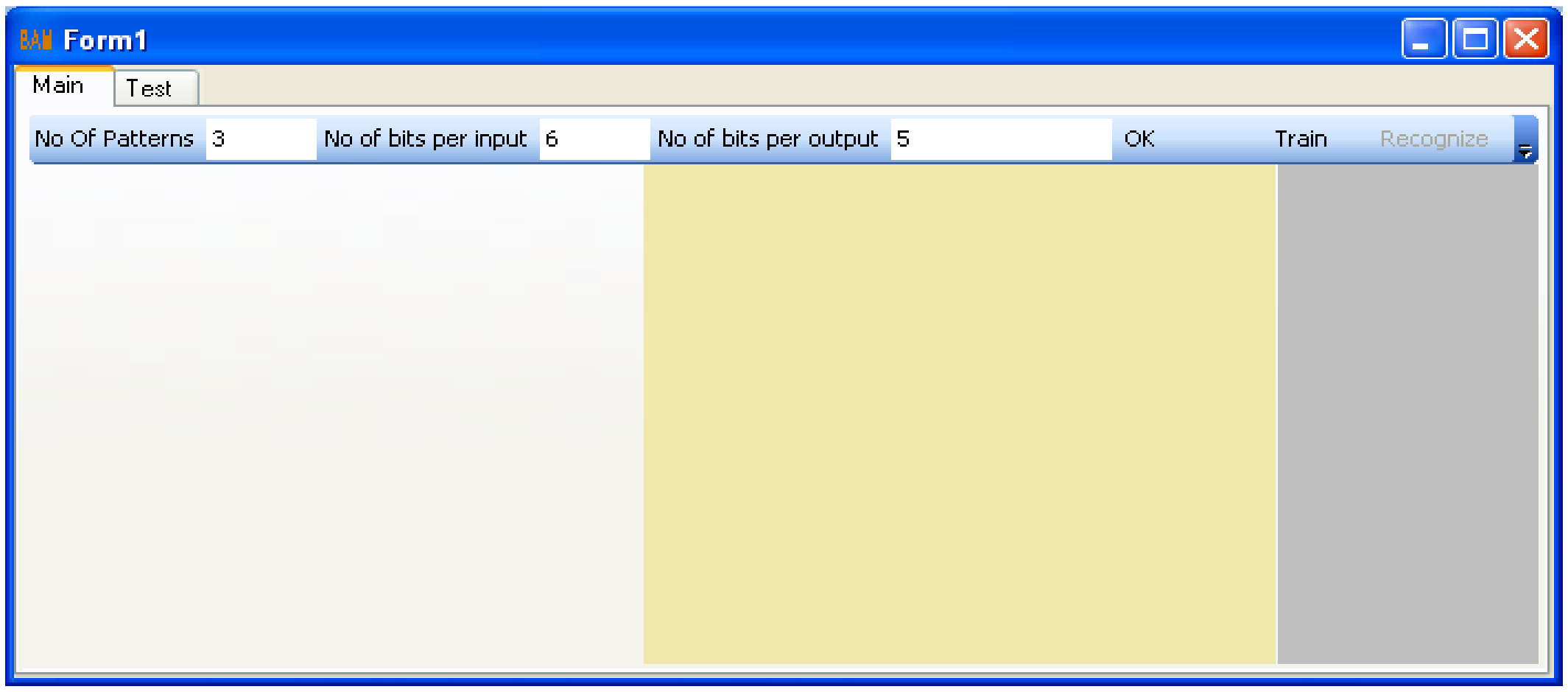

Figure 4 Screen showing how to setup network

In the figure 4 *No of Patterns* enumerates the number of patterns we need to use in training, *No Of Bits* Per Input specifies number of bits desired to use for each input and *No Of Bits* Per Output stipulates number of bits to use for each output.

Once the required information has given, OK button has to be pressed in the application which provides enough fields to enter input and output pairs.

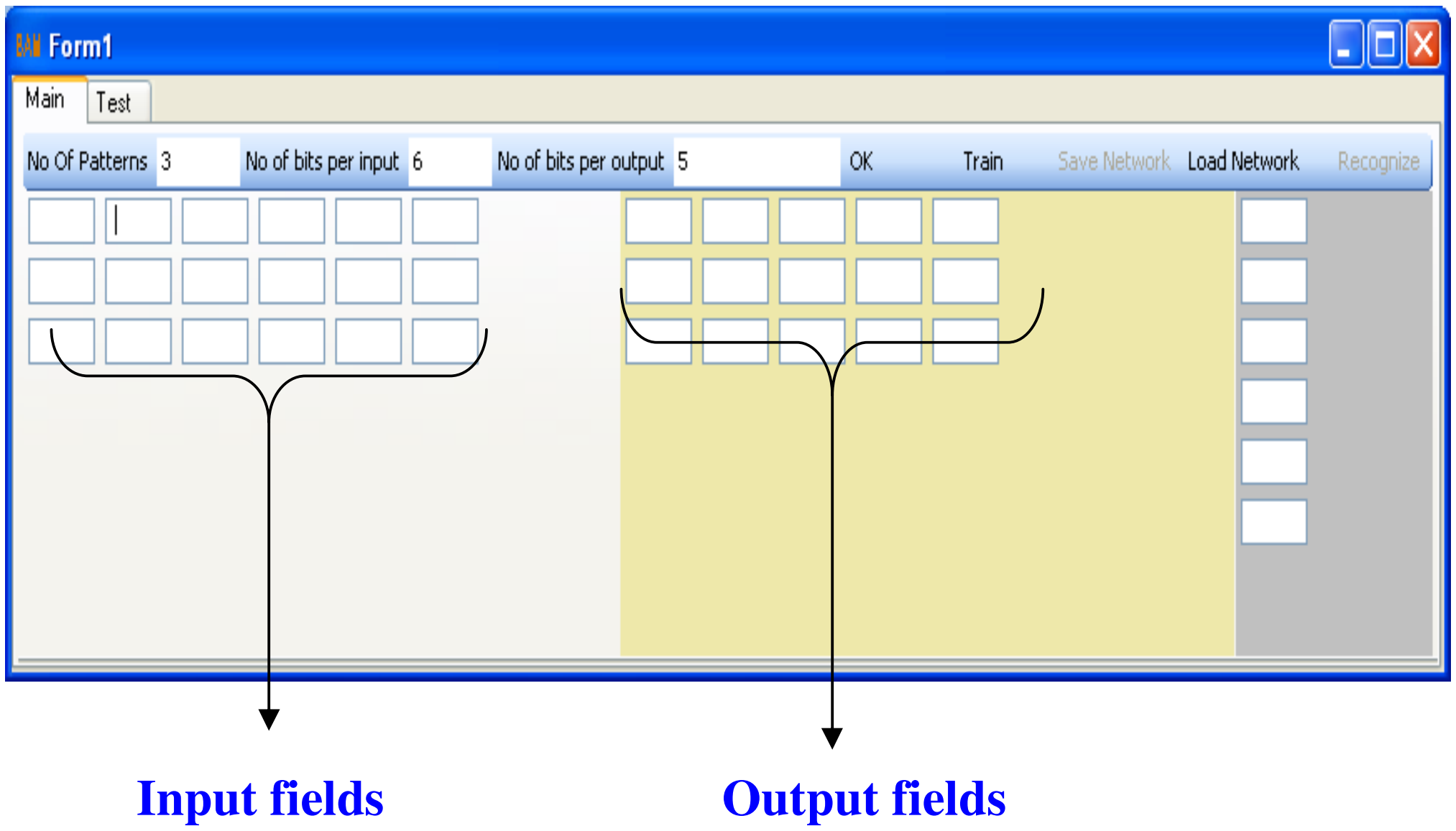

Figure 5 Screen showing how to give input for BAM

## TRAINING THE NETWORK:

Once the required training set has given to the above process press *Train* button to make the application to undergo training process .

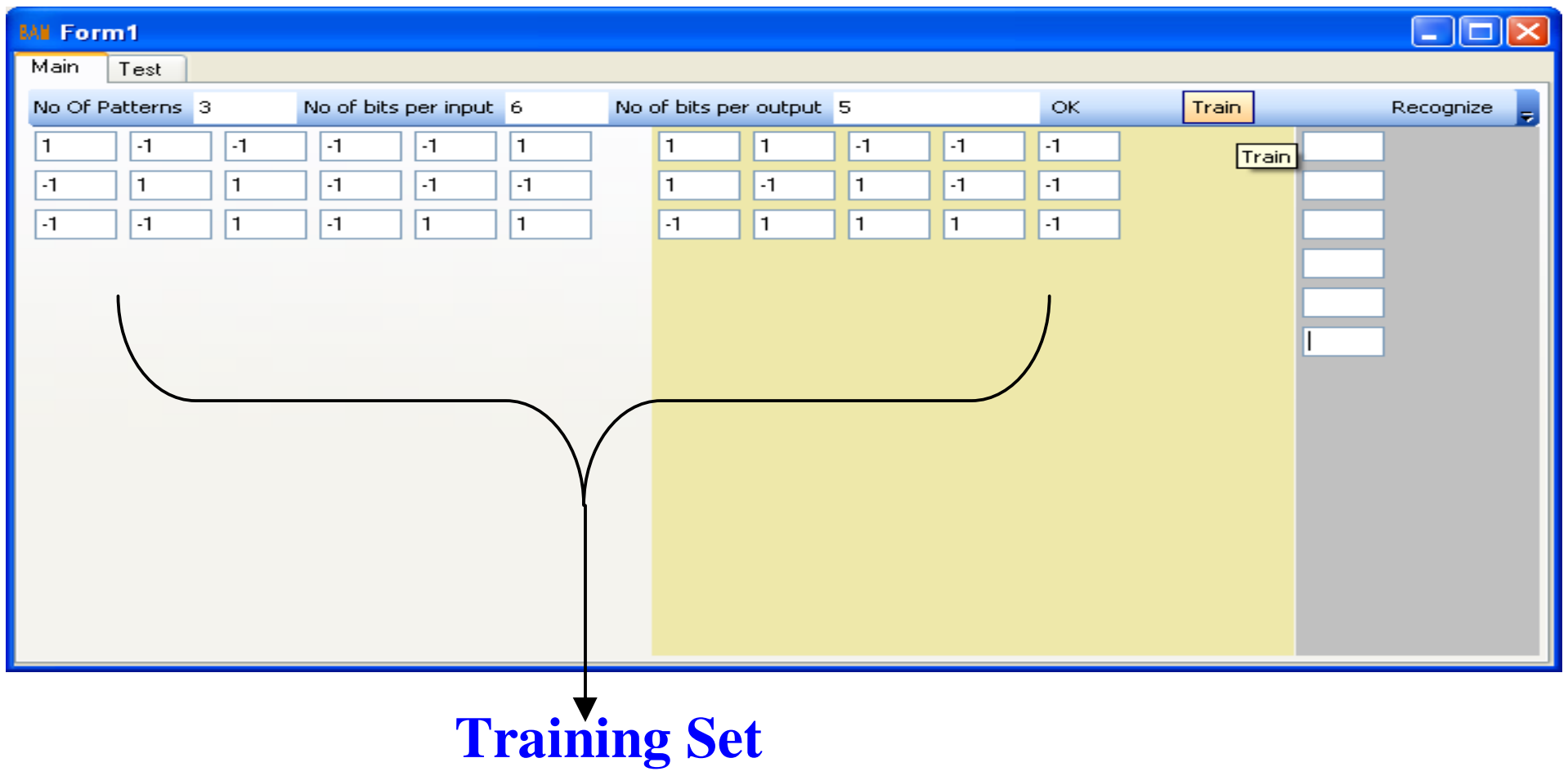

Figure 6 Screen showing how to take Training Set for BAM

Afterthe training is completed it will be manifested weight matrix.

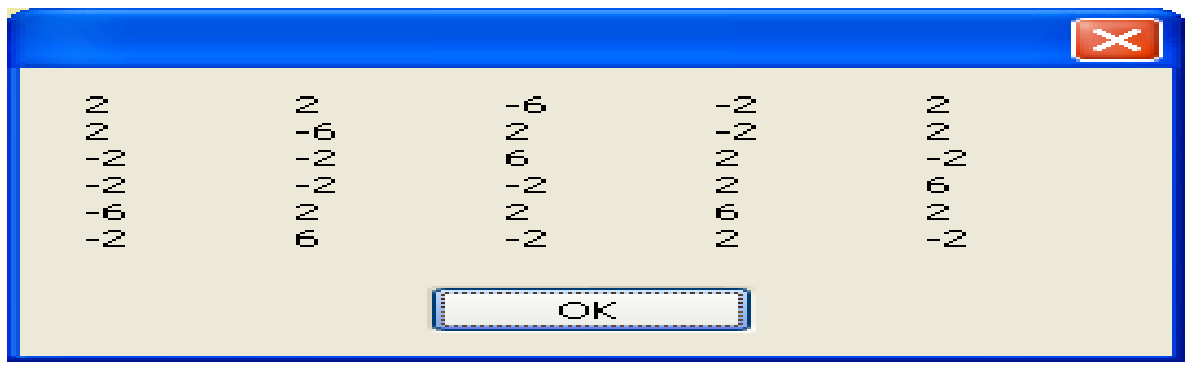

Figure 7 Screen showing Completion of BAM Training

## USER AUTHENTICATION USING BAM

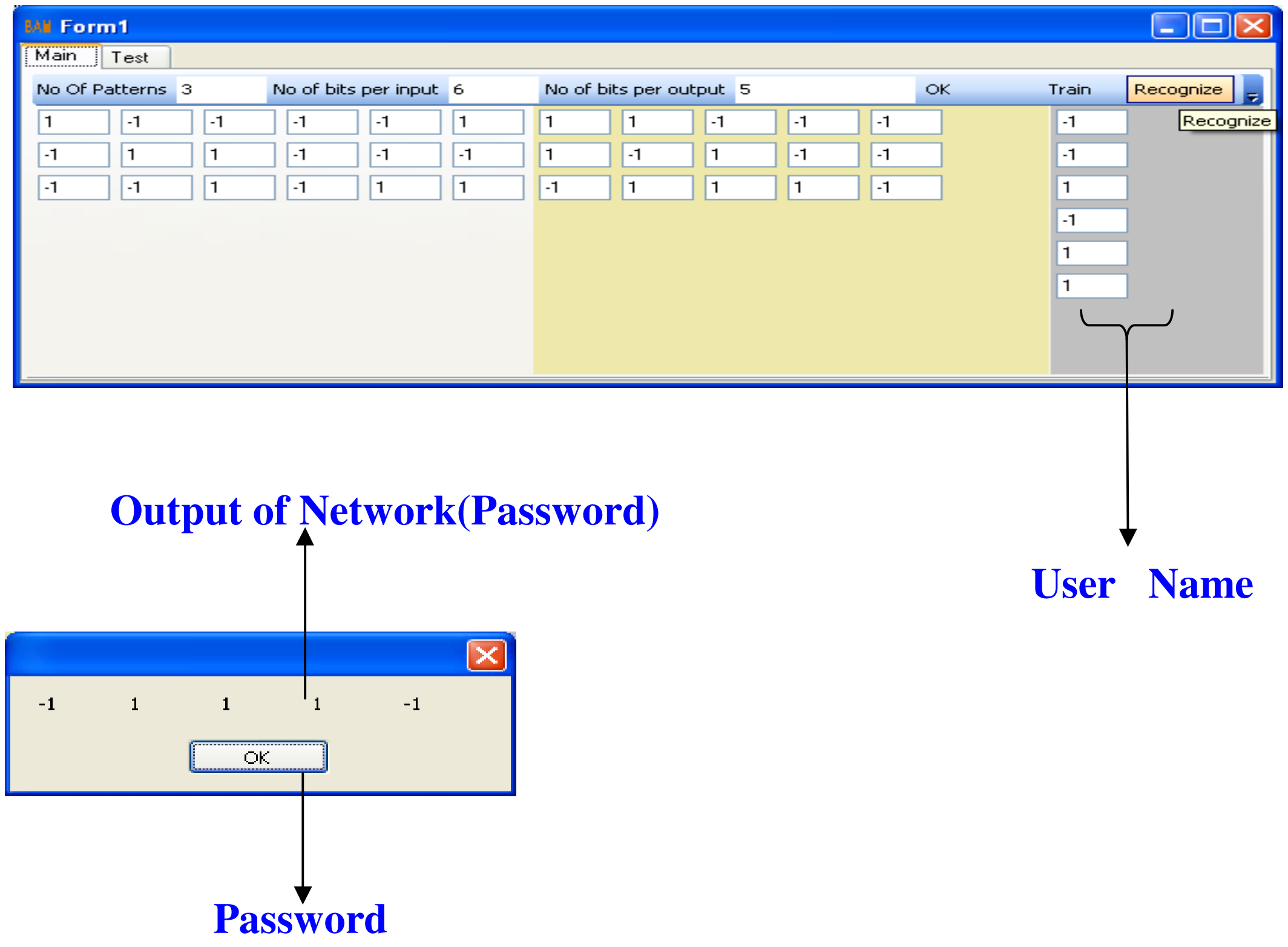

Figure 8 Screen showing Completion of BAM Authentication

Here a comparison can be done to the output of the network with the password given by the user and, if both are the same then the user is a valid user which gives him an opportunity to get serviced.

## BAM FOR GRAPHICAL PASSWORDS:

Entering image as password for any applicationaheich uses BAM network is impossible. So the application has to convert the given image in to binary form.

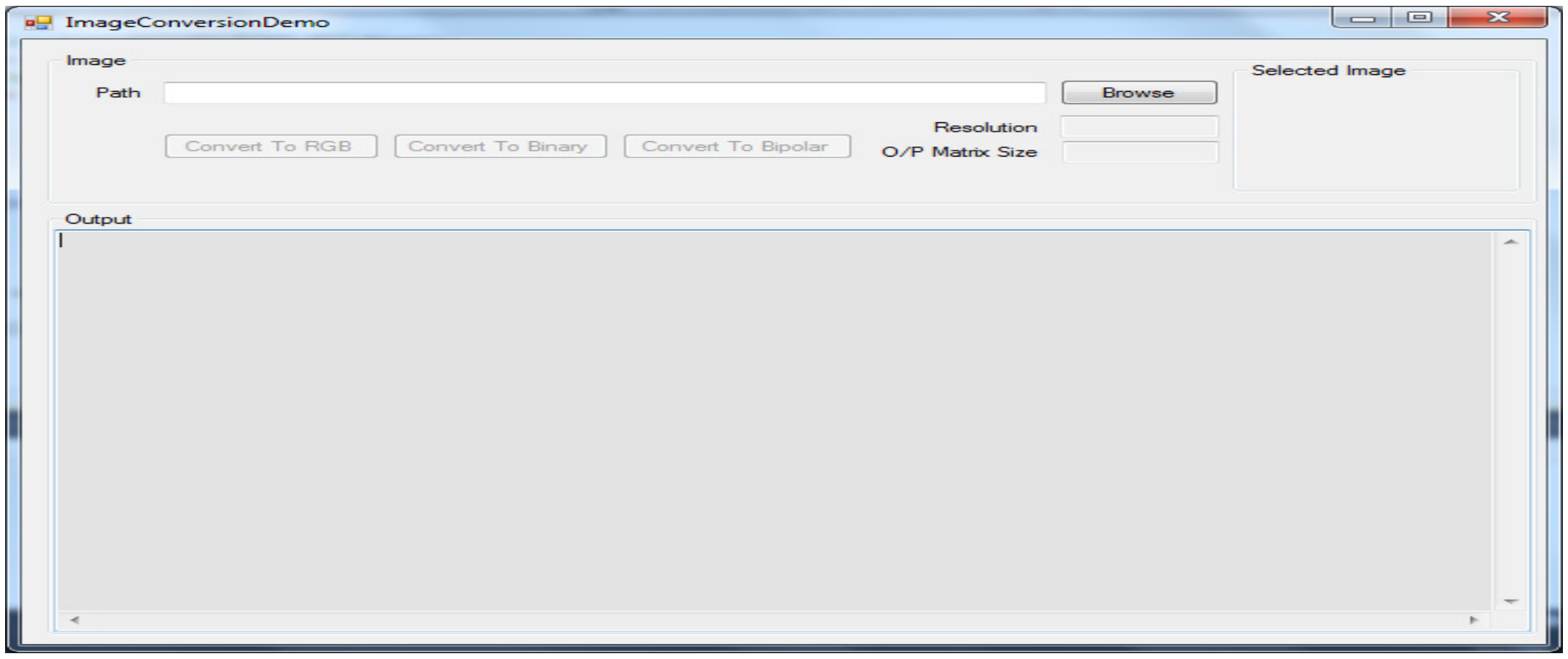

Figure 9 Screen showing how to select an the image

In figure 9 the Path cannotes path of the image we want to convert, Selected Image shows the selected image ,Resolution denotes the resolution of the selected image and O/P Matrix Size specifies size of the output matrix. After an image has been selected by the application it will be displayed in the Selected Image box and conversion buttons will be provided to convert image to different forms. The aplictaion selects an imageas an input from any one of the systems .Once the image is selecetd y the application it will be displayed as a thumb nail and resoultion values of the image is also dispalyed as shown in figure 10.

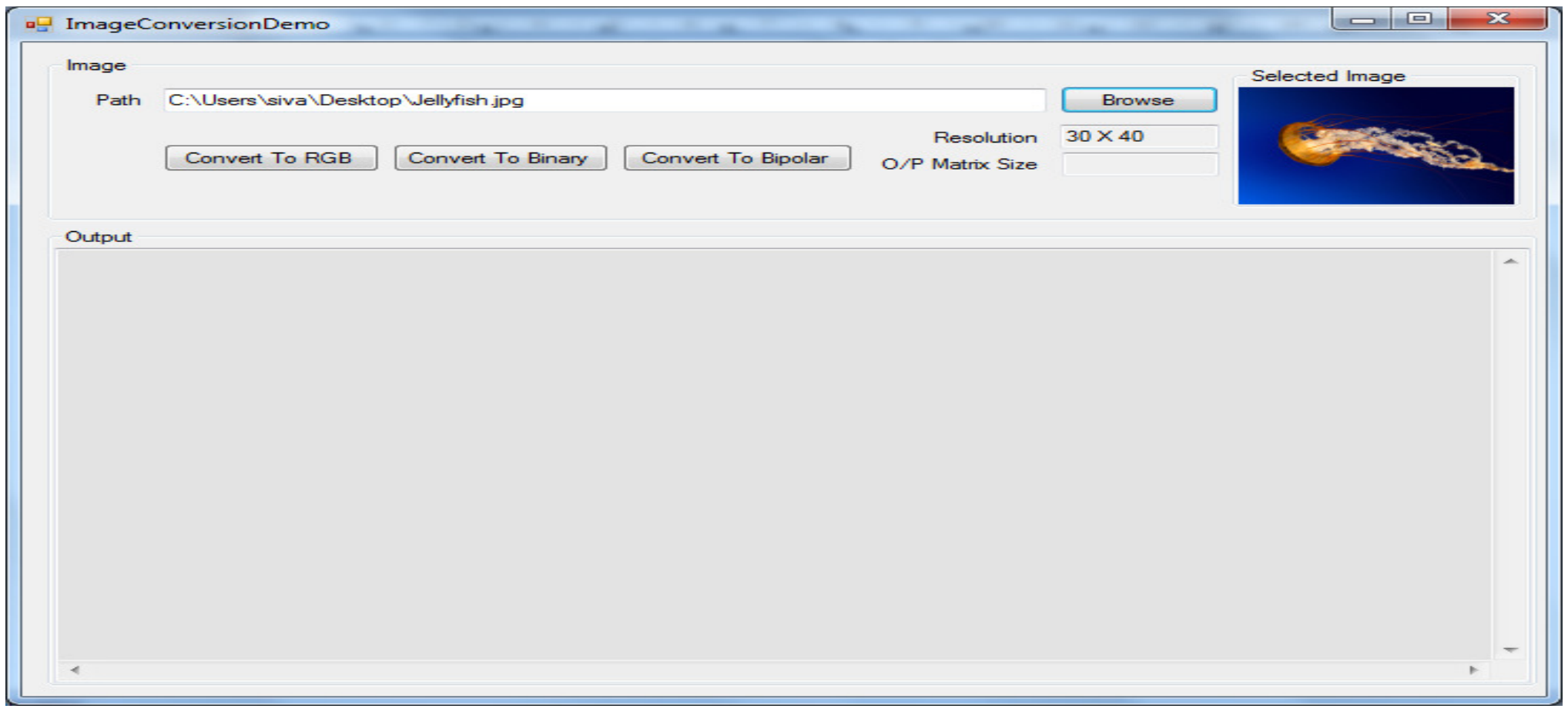

Figure 10 Screen showing how to convert an image into an Integer (RGB) matrix

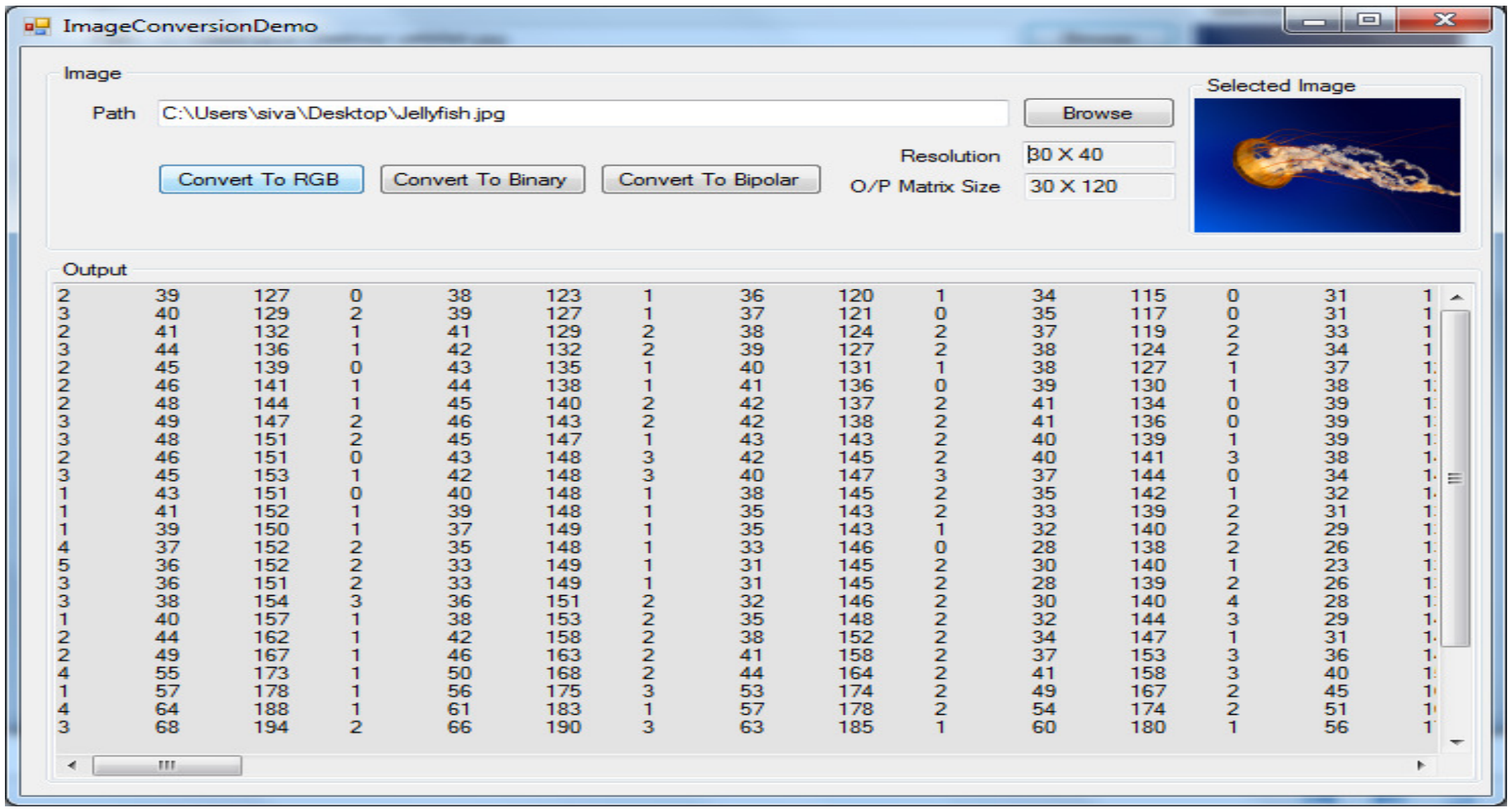

Figure 11 output matrix

After electing an image “Convert To RGB“is pressed then we can monitor the output matrix in output box in figure 11 and also it displays the size of the output matrix. The sceern shows the integer values of an image.

These integer values will be converted in to binary values so as to satisfying our probablistic method,wher the input to the network should be normalized values .

After selecting an image if “Convert To Binary“button is pressed then we can scrutinize the output matrix in the output box as shown in figure 12.

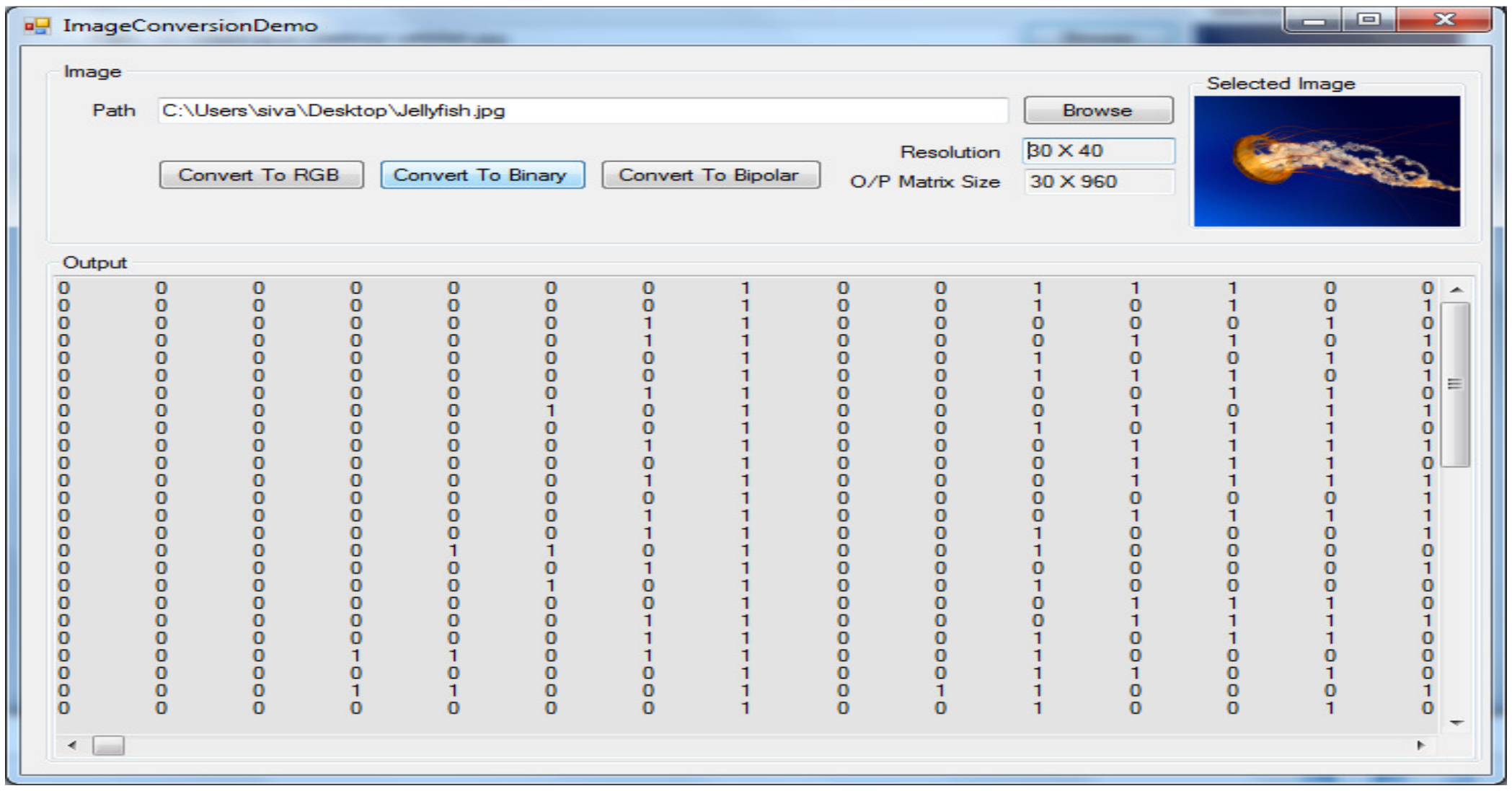

Figure 12 Screen showing how to convert an image into a binary matrix

This screen shows all the binary values corresponding to the inateger values of the image which has been selected through the application. Thes binary values can be used for the trainig the Hopfiled network for graphical password authentication

The application also converts the binary values in to Bipolar values after pressing the button” “Convert To Bipolar“. Then we can perceive output matrix in output box.

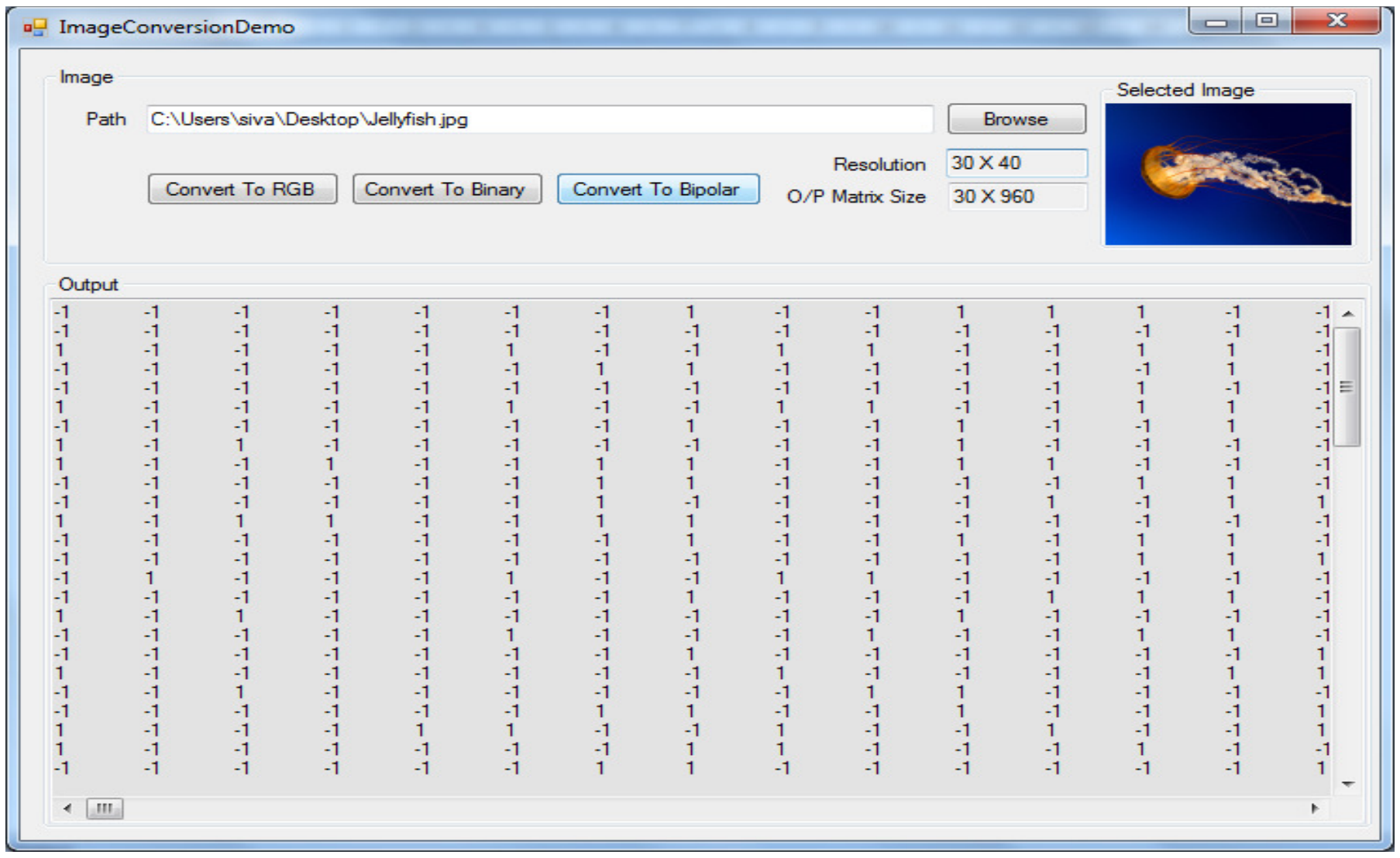

Figure 13 Screen showing how to convert an image into a bipolar matrix

In the above classification we have seen how to convert an image into text. Once the image has been converted we can use it as normal password

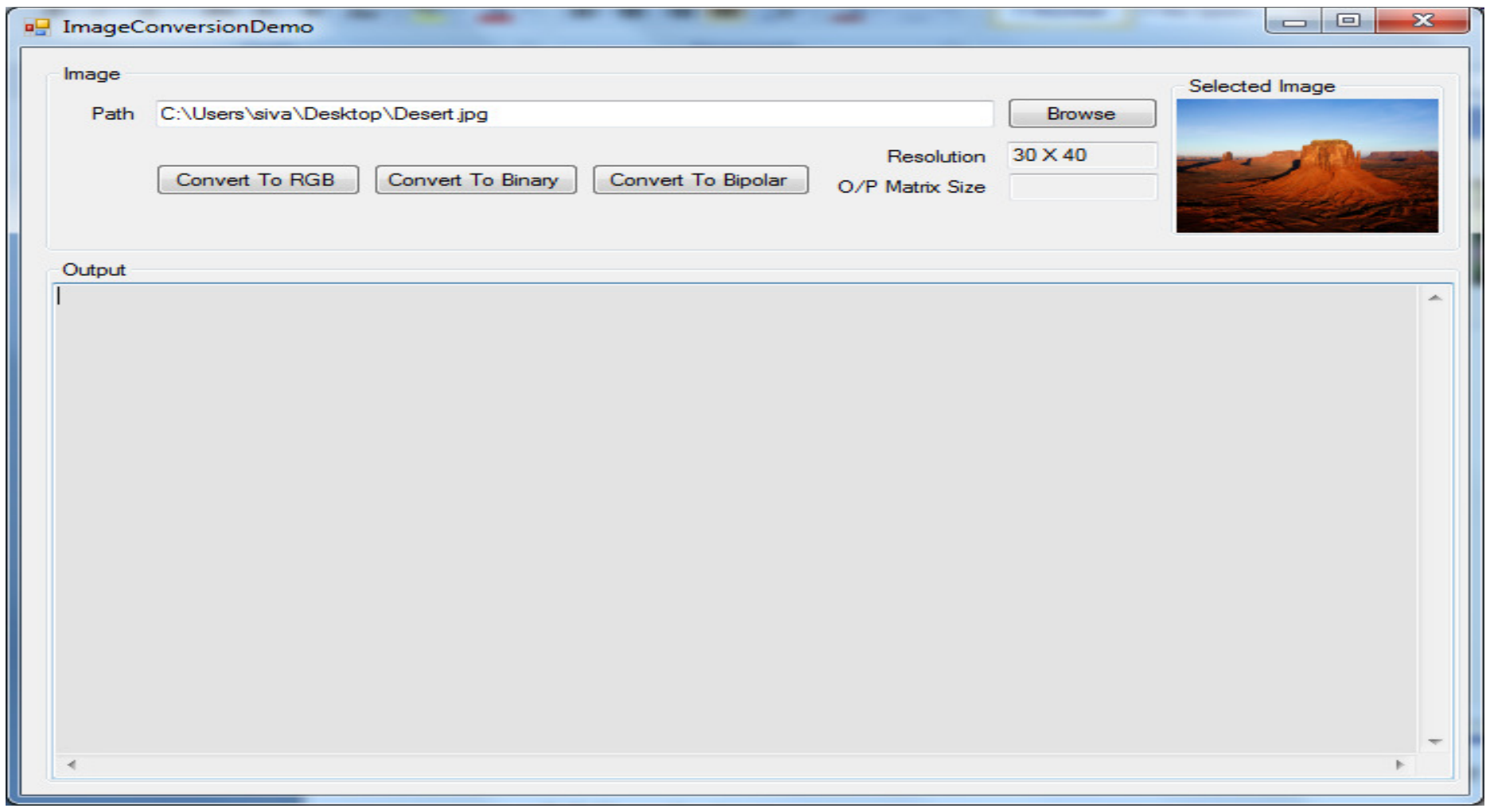

Figure 14 Screen showing selecting an image

# 5 CONCLUSIONS:

In this paper probalistic approach for password authentication using one of the associative memory BAM is introduced.results show that this method can overcome the limitations of Conventional Password Authentication Scheme.In this approach of password authentication using BAM we need not give input and output together to train the network.Since BAM is bidirectional in nature we can further improve this method for an application which can give username if the password is given. But this method may have some limitations that differnt usernames may have the same password.Inorder to solve this we can take password plus any unique image as input for identifying the username, if the same combination was given as input while training network.

## ACKNOWLEDGEMENTS

The authors wish to thank all the referees for their valuable suggestions to improve this paper.

## Authors

†A.S.N Chakravarthy received his M.Tech (CSE) from JNTU, Anantapur , Andhra Pradesh, India. Presently he is working as an Associate Professor in Dept. Of Computer Science and Engineering in Sri Aditya Engineering College, SuramPalem, E.G.Dist, AP, India. His research areas include Network Security, Cryptography, Intrusion Detection, Neural networks.

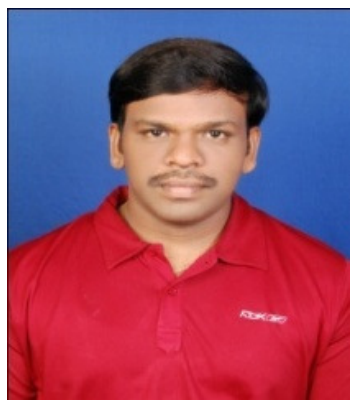

†† Penmetsa V Krishna Raja received his M.Tech (CST) from A.U, Visakhapatnam, Andhra Pradesh, India. He is a research scholar under the supervision of Prof.P.s.Avadhani. His research areas include Network Security, Cryptography, Intrusion Detection, Neural networks.

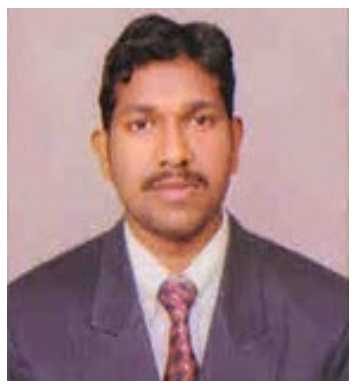

†††Prof. P.S.Avadhani did his Masters Degree and PhD from IIT, Kanpur. He is presently working as Professor in Dept. of Computer Science and Systems Engineering in Andhra University college of Engg., in Visakhapatnam. He has more than 50 papers published in various National / International journals and conferences. His research areas include Cryptography, Data Security, Algorithms, and Computer Graphics, Digital Forensics and Cyber Security.

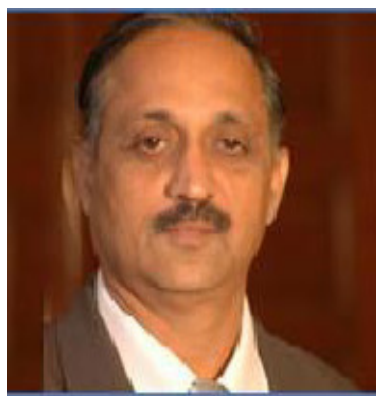

o0o